# Probing Spin Accumulation induced Magnetocapacitance in a Single Electron Transistor

T. H. Lee[1,2,3] and C. D. Chen[3]

[1]*Department of Physics, National Taiwan University, Taipei 106, Taiwan*

[2]*Nano Science and Technology Program, Taiwan International Graduate Program, Academia Sinica, Taipei 115, Taiwan*

[3]*Institute of Physics, Academia Sinica, Taipei 115, Taiwan*

**The interplay between spin and charge in solids is currently among the most discussed topics in condensed matter physics. Such interplay gives rise to magneto-electric coupling, which in the case of solids was named magneto-electric effect, as predicted by Curie on the basis of symmetry considerations. This effect enables the manipulation of magnetization using electrical field or, conversely, the manipulation of electrical polarization by magnetic field. The latter is known as the magnetocapacitance effect. Here, we show that non-equilibrium spin accumulation can induce tunnel magnetocapacitance through the formation of a tiny charge dipole. This dipole can effectively give rise to an additional serial capacitance, which represents an extra charging energy that the tunneling electrons would encounter. In the sequential tunneling regime, this extra energy can be understood as the energy required for a single spin to flip. A ferromagnetic single-electron-transistor with tunable magnetic configuration is utilized to demonstrate the proposed mechanism. It is found that the extra threshold energy is experienced only by electrons entering the islands, bringing about asymmetry in the measured Coulomb diamond. This asymmetry is an unambiguous evidence of spin accumulation induced tunnel magnetocapacitance, and the measured magnetocapacitance value is as high as 40%.**

Correspondence and requests for materials should be addressed to C.D.C. (chiidong@phys.sinica.edu.tw)

**Introduction**

The study on magnetocapacitance has been motivated by its fundamental interest and high practical importance, and that stimulated the revival of attention in this field about one decade ago[1]. The potential applications include magnetic-field sensors[2] and electric-write magnetic-read memory devices[3]. Experimentally, two types of magnetocapacitance were revealed: The first type is magnetic-field strength dependent, and is typically found in multiferroic materials with perovskite superlattice structure such as $BiMnO_3$[4] and $La_{0.875}Sr_{0.125}MnO_3$[5]. The second type, also known as tunnel magnetocapacitance (TMC), is found in magnetic tunnel junctions consisting of $AlO_x$[6,7] and $MgO$[8] barriers. For the latter, the magnetocapacitance varies with magnetization alignment configuration of the two constituent ferromagnetic electrodes, and the charge screening length at the interface of the tunnel barrier was predicted to be elongated by the spin-dependent diffusion constants[9,10]. This extended screening length was then used to deduce the effective capacitance that departed from the geometrical capacitance. However, this simple approach may overlook the details of the charge distribution, which can be important for the determination of effective capacitance value. Therefore, a precise calculation of charge distribution on microscopic scale is desirable. Furthermore, because the screening length cannot be directly measured, the physical picture of the microscopic origin behind TMC is hardly appreciated.

Here, we present a clear microscopic mechanism for TMC. When magnetic tunnel junction is aligned in anti-parallel (AP), accumulation of minority spins and depletion of majority spins take place at the interfaces[11,12], and this induces a difference in the interfacial Fermi levels of the majority and minority spins[13]. Thus, the spins diffuse from the interface with different diffusion lengths, and that gives rise to a difference between majority/minority spin density distributions. As a result, there are distinct, adjacent accumulation zone and depletion zone in the total charge density distribution that form a tiny charge dipole. This charge dipole, rather than single exponential charge decay, corresponds to an extra serial capacitance that is responsible for the measured TMC. The extra serial capacitance poses an extra energy required for spins to flip when electrons tunnel through an AP aligned magnetic junction, giving rise to the observed TMC.

To determine junction capacitance, AC impedance method is typically used. However it may involve complex frequency-dependent dispersion behavior[14]. In our study, we get rid of this complication by incorporating magnetic tunnel junction into a single-electron-transistor (SET). The junction capacitance is accurately determined by utilizing the charging effect. Specifically, the junction capacitance in various magnetization alignment configurations were derived from the slopes of the Coulomb diamond[15]. With this, electrons that tunnel into the island through an AP-aligned junction always experience TMC. This TMC is unambiguously confirmed by the observed asymmetric Coulomb diamond in an anti-parallel (AP) aligned ferromagnetic single-electron-transistor. In the sequential tunneling regime, a single spin flip costs an extra energy originated from the TMC.

**Methods & Results**

To observe extra energy cost by single spin flip, a ferromagnetic single-electron-transistor was designed and constructed. However, the spin diffusion length in ferromagnetic materials is generally too short to sustain spin accumulation[16]. In order to appreciate the magnetocapacitance effect, we used a thin nonmagnetic layer to cover one of the ferromagnetic electrodes. In this way, stable charge dipole is generated inside the nonmagnetic layer. The device, as schematically shown in Figure 1a, was fabricated using the standard electron-beam lithography technique and two-angle evaporation method. The source and drain electrodes were made of Co, whereas the island was made of 10nm-thick permalloy (Py, $Ni_{80\%Wt}Fe_{20\%Wt}$), a ferromagnetic material with a smaller coercivity. The Py island was directly covered with a 2nm-thick aluminum (Al) layer and then a thin tunnel barrier, which was formed by direct evaporation of alumina ($Al_2O_3$) crystal without oxidation process. Since the Al layer is much thinner than ferromagnetic Py layer, its superconductivity is suppressed by the proximity effect[17-20]. As illustrated in Figure 1b, the device consists of $Co/Al_2O_3/Al/Py/Al/Al_2O_3/Co$ with two tunnel junctions in series. The junction area $A$ measured about 65nm × 65nm, corresponding to a charging energy of the order of 6K. In addition, a gate-electrode was located about 800nm away from the island, giving a $C_g$ value of about 0.4aF and a Coulomb oscillation period of about 0.5V. All $I$-$V_b$ characteristics were measured using 4-point probe technique at 120mK.

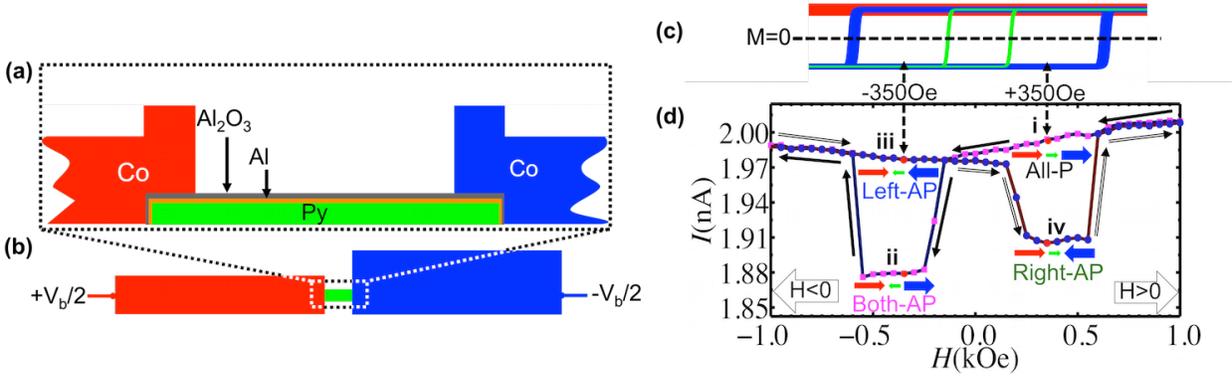

**Figure 1 (color online) | Fabrication and measurement process.** (a) and (b) schematic drawings of the device in cross-section view and top view. Throughout this paper, we use blue, green, and red to indicate right electrode, center island, and left electrode, respectively. The magnetic field is applied parallel to the long-axis of the two electrodes. The right electrode is made wider to decreases the coercivity. (c) Magnetization curves for right electrode (blue), center island (green), and left electrode (red). Note that the curve for left electrode shows no field-dependence because of the large coercivity. (d) Tunnel magneto-current measured at $V_b$=8.7mV, a bias much greater than the maximum Coulomb blockade threshold voltage of $2E_C/e \approx 1$mV, so that the current is gate-independent and the TMR effect can be clearly observed. The magnetic field is swept (indicated by black arrows along the curve) sequentially from *i* to *iv*, with magnetization directions indicated by blue (right electrode), green (center island), and red (left electrode) arrows.

The device was symmetrically voltage biased at $+V_b/2$ and $-V_b/2$ on the left and right electrodes, respectively. The magnetic field was applied along the long axis of the cobalt electrodes such that the electrode and island could either be aligned in parallel (P) or AP configuration. Prior to the measurement, the device was subjected to a magnetic field of +50kOe, to ensure that the magnetization directions of both left and right electrodes and the center island were all parallel to the external field. At this stage, analysis on the conductance height along the edges of Coulomb diamond indicated that the resistance ratio between the left and right junctions ($R_{left}/R_{right}$) was approximately 1.47. Manipulation of magnetization was then carried out by ramping the magnetic field between ± 1000Oe.

The width of the left electrode was deliberately designed to be narrower so that the magnetization of this electrode remained unchanged due to large shape anisotropy and only magnetizations in the island and right electrode changed direction within this ramping field range. The expected magnetization curve in this field range is illustrated in Figure 1c, and the measured magneto-current trace depicted in Figure 1d displays only a single step of suppression in both ramping up and down directions. The plateau appears between 250Oe and 550Oe and hence the alignment configuration is well defined and stables at $H_\pm = \pm350$Oe. In terms of alignment configurations between the island and electrodes, there are in total four configurations, which are referred to as All-P, Left-AP, Right-AP, and Both-AP. A careful analysis on the $I$-$V_b$ characteristics at various gate voltages in these four configurations revealed that the tunneling conductance of both junctions in the AP configuration is reduced by a factor of 1.62 as compared to the P configuration. This TMR effect, together with the difference in left and right junction resistance, results in different current amplitudes in these four configurations. Based on the current values at $H_\pm$, we identified the alignment configuration from the largest current as All-P, Left-AP, Right-AP, and Both-AP.

We then took a pause at $H_\pm$ in each ramping direction and measured a batch of $I$-$V_b$ characteristics at varying $V_g$. The charging energy $E_C$ of the device can be deduced directly from the stability diagram (*i.e.* Coulomb blockade diamond) measured in All-P configuration because in this configuration the device can be viewed as a usual non-magnetic SET. The diamond is symmetric in respect to $V_b$=0 and $n_g \equiv C_g V_g/e$=integer point. The $E_C$ is determined to be 510μ$e$V from the triangular area in both bias polarities. By computing the slopes of the diamond borders, the junction capacitances are determined to be $C_{Left}$ = 101.8aF and $C_{Right}$ = 55.1aF for the left and right junctions, respectively.

Clear stability diagrams were also obtained for the other three configurations. It is interesting to note that the diamond can become asymmetric in these configurations. To clearly depict the deviation in each alignment configuration, four diamonds are stacked altogether in Figure 2a. The four borders of these diamonds are indicated by $L_{in}$, $R_{out}$, $L_{out}$, and $R_{in}$. *L* and *R* stand for tunneling taking place at the left and right junctions, respectively; while the subscription denotes the electron tunneling direction in respect to the island. In the Right-AP configuration the $L_{out}$ border tilts clockwise, while in the Left-AP configuration the $R_{out}$ border tilts anti-clockwise. In Both-AP, the two borders tilts to overlap with both the $L_{out}$ of Right-AP and the $R_{out}$ of Left-AP. The change in the border slopes in these three configurations signifies a

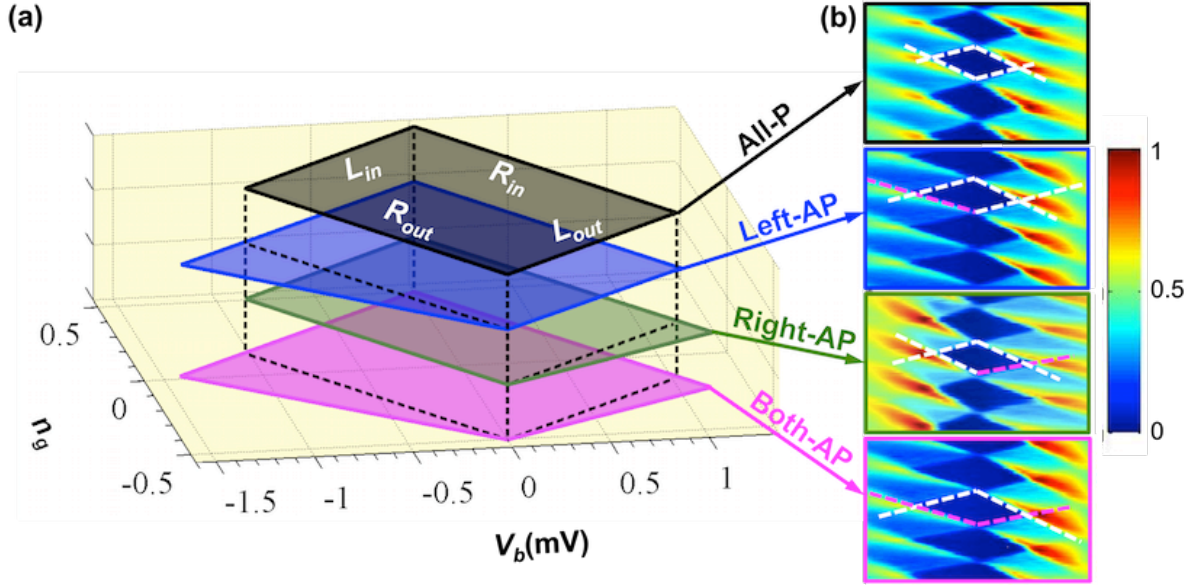

**Figure 2 (color online) | Measured Coulomb diamonds in 4 different alignment configurations.** The diamonds are presented in color intensity plots in (b) and are stacked together for comparison in (a). In (b), the dashed lines mark the borders of Coulomb diamonds, and their slopes are used to evaluate effective junction capacitances.

change in the junction capacitance. We recall that, in an SET, the threshold for electrons to tunnel, *i.e.* the border, through a junction is determined by the capacitance of the counterpart junction. A smaller capacitance corresponds to a more outwardly tilted border. Hence, our observations in Figure 2 can be summarized as follows: when tunneling *out* from the island through a junction, the electrons experience a decrease in the capacitance of the counterpart junction, *if and only if* the counterpart junction is in AP configuration. This condition of a decrease in the capacitance seen by the tunneling electrons is referred to as the TMC criterion.

In order to characterize quantitatively the diamond asymmetry, we introduce an effective capacitance of Co/Al$_2$O$_3$/Al/Py, $C_{eff} = \left[1/C_{Co} + 1/C_0 + 1/C_{Al} + 1/C_S\right]^{-1}$. $C_0$ is simply the intrinsic geometrical capacitance of Al$_2$O$_3$, given by $C_0 = \Delta Q/\Delta V$, where $\Delta V$ is the electrical potential difference applied across Al$_2$O$_3$ and $\Delta Q$ is the charge induced inside Al$_2$O$_3$. Note that $\Delta Q$ is held inside Al$_2$O$_3$ by an *E*-field given by $-\Delta V/d$, where $d$ is the thickness of Al$_2$O$_3$. This *E*-field, however, leaks through both Co/Al$_2$O$_3$ and Al$_2$O$_3$/Al interfaces, inducing screen charge density $e\Delta n(x)$, which in turn damps the leakage *E*-field. As a result, both $e\Delta n(x)$ and $V(x)$ decay exponentially away from interfaces with a characteristic screening length $\xi$ of typically 0.5–1Å[21].

Despite this decay, the ratio between $e\Delta n(x)$ and $dV/dx$, remains a constant inside Co and Al. This way, $e\Delta n A dx/dV$ inside Co and Al can be, respectively, viewed as 2 serial capacitances, $C_{Co}$ and $C_{Al}$, adjacent to $C_0$. These interfacial capacitances, $C_{Co}$ and $C_{Al}$, depend only on their respective electron densities. Lastly, a spin capacitance $C_S$ originated from the TMC effect is used to account for the difference between All-P and other three AP-configurations. To quantify this effect, a TMC ratio, defined as $\Delta_{TMC}=C_{eff}/C_S$, is introduced. Since $C_{eff}$ is simply the product of electron charge and slopes of Coulomb diamond borders, it can be readily obtained for each border in all alignment configurations. We further set $\Delta_{TMC}$ to zero in All-P case, so that $\Delta_{TMC}$ in the three AP-configurations can be evaluated, and the values are listed in Table 1. From this table, we clearly observe that the junction exhibits substantial $\Delta_{TMC}$ whenever the magnetocapacitance criterion is fulfilled. The remaining is negligibly small within the error bar. The observed $\Delta_{TMC}$ value of approximately 40% is the highest among $AlO_x$-based magnetic tunnel junctions.

**Discussion**

For illustration purpose, we assume that the majority spin in Al is up, denoted by (+) sign, and vice versa. Then, we derive $C_S$ using spin-dependent drift-diffusion model. The detail of this model is provided in the Supplementary Information (S1-S3). It is merely a steady solution to the equation of motion for spin-dependent drifting electrons and back-diffusion electrons inside Al: $-\sigma_\pm(x)\partial^2 V_\pm(x)/e\partial x^2 = D_\pm(x)\partial^2 n_\pm(x)/\partial x^2$, where $x$ is the distance from the $Al_2O_3$/Al interface. The drifting and back-diffusion processes are described by the left-hand side and right-hand side of this equation, respectively. $\sigma_\pm$ is conductivity, and $D_\pm$ is diffusion coefficient. The chemical potentials $V_\pm(x)$ satisfy the spin-dependent Poisson equation, and $V_+ - V_-$ obeys the diffusion-relaxation equation[11], as described in S1.

Our task is to get a solution for the charge potential $V(x)=\left(n_+(x)V_+(x)+n_-(x)V_-(x)\right)/\left(n_+(x)+n_-(x)\right)$ and total charge density perturbation $\Delta n(x)=n_+(x)+n_-(x)-n_0$; where $n_\pm(x)$ are spin-dependent electron densities, and $n_0$ is the initial uniform electron density of Al. In All-P configuration, there is no spin accumulation, and the calculated interfacial capacitance is simply the one that incorporates the charge screening at the interface, i.e. $C_{Al} \equiv dQ(x)/dV(x) = e\Delta n(x)Adx/dV(x)$. The calculated $C_{Al}$ is an $x$-independent constant and can serve as a baseline for the calculations of extra serial $C_S$ in the AP-configurations.

In AP-configurations, the current flowing through Al for each spin is not steady (Figure 3a), giving rise to spin accumulation. This accumulation, in turn, causes a difference between spin-up and spin-down diffusion lengths, $\lambda_{\pm}$. This difference is taken into account in our finite element analysis described in S2, yielding a solution of $\Delta n(x)$ different from that of All-P case. This $\Delta n(x)$ is simply comprised of two spin-dependent components (Figure 3b), i.e., $\Delta n = (\Delta n_{C,+} + \Delta n_{S,+}) + (\Delta n_{C,-} + \Delta n_{S,-})$, where $\Delta n_{C,\pm}(x) \propto \exp(-x/\xi)$ is due to $E$-field penetration alone, which becomes negligible for $x \gg \xi$, while $\Delta n_{S,\pm}(x) \propto \exp(-x/\lambda_{\pm})$ is due to spin-accumulation in AP-configuration. Note that each accumulated/depleted spin diffuses with different spin diffusion length, in contrary to spin-independent charge screening length $\xi$. This difference in $\lambda$ is responsible for the location of $x_c$, where $\Delta n_{S,+}$ and $\Delta n_{S,-}$ cancel each other out, forming a tiny charge dipole structure. In other words, $x_c$ is just the solution of $\Delta n_{S,+}(x = x_a)\exp(-x_c/\lambda_+) + \Delta n_{S,-}(x = x_a)\exp(-x_c/\lambda_-) = 0$, which serves as a critical point across where $\Delta n(x)$ changes sign. In the limit of small current that occurs within Coulomb blockade regime, the difference between $\lambda_{\pm}$ is small such that the solution for $x_c$ fall within Al thickness, allowing the charge dipole to exist inside Al with $\Delta n(x)=0$ at $x=x_c$. Because of the finite $\Delta n(x)$ value at $x>x_c$, an extra serial capacitance $C_S$ in AP-configuration is generated, which can be calculated using $1/C_{Al} + 1/C_S = dV(x)/e\Delta n(x)Adx$, as shown in Figure 3c. Note that the calculated $1/C_S$ is only dependent on the magnetic configuration of our device since $1/C_{Al}$ is used as a reference for $dV(x)/e\Delta n(x)Adx$ calculation. The induced tiny charge dipole, having an equivalent capacitance $C_S$, thus acts as a serial capacitance to $C_{Al}$ in the AP-configuration. The existence of this charge dipole requires that $x_c$ be greater but close to the screening length $\xi$. While the lower limit of Al thickness is set by $x_c$, the upper limit is set by the spin diffusion length $\lambda$ and the effect of exchange proximity (see S1). Since it is a single electron sequential tunneling process, this implies an extra cost of "charging energy" for a single spin flip event,

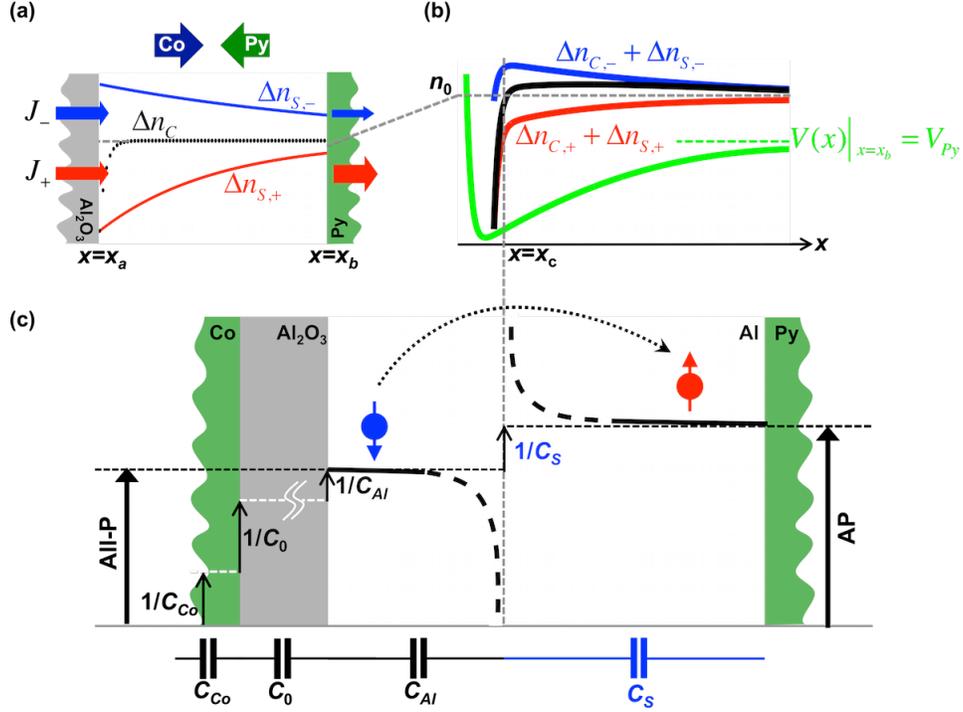

**Figure 3 (color online) | Illustration of charge and spin distributions in the AP-configuration.** (a) Spin accumulation (and depletion). Blue and green arrows indicate magnetization direction of Co and Py, respectively. Current density $J$ is composed of spin up (red) and spin down (blue) components, represented by corresponding colored arrows, whose width indicates the relative magnitude. Dotted black curve resembles the charge density decay in All-P configuration (Figure S1c, S1), and serves as a baseline. Smooth red and blue curves depict spin up depletion and spin down accumulation, respectively. Note that the areas enclosed between the red/blue curve and the equilibrium level $n_0$ (gray dashed line) are the same. (b) A blow-up view of $n(x)$ and $V(x)$ around $x_c$, at which $n(x)$ crosses $n_0$. Perturbation in $n(x)$ (black curve) is composed of the spin up (red curve) and spin down (blue curve) contributions. The green curve shows charge potential $V(x)$ with the block dotted line indicating the height of $V(x)|_{x=x_b} = V_{Py}$. (c) Deduction of inversed capacitance values from $dV(x)/e\Delta n(x)Adx$. All-P plateau (black dash line, left) is elevated to AP plateau (black dash line, right) for $1/C_S$ after $x_c$. The arrowed sphere (blue and red) represents a single electron spin right after experiencing sequential tunneling from tunnel barrier $Al_2O_3$, whereas the black dotted arrow indicates the flipping event at the cost of $e^2/2C_S$.

which takes place in AP-configuration. The positive side of the charge dipole is close to the $Al_2O_3$ interface, regardless of the current direction, as explained in S3. However, if the electrons flow in opposite direction to the one shown in Figure 3a, *i.e.* from Py island towards $Al_2O_3$/Al interface, the charge accumulates from $Al_2O_3$/Al interface and compensates the positive side of the charge dipole. This would wash away the charge dipole, and the TMC diminishes, causing Coulomb diamond to display asymmetry with respect to the bias direction of $V_b$.

To conclude, we observed and verified the TMC effect in a ferromagnetic SET based on capacitance values determined from the asymmetry of Coulomb diamonds. The asymmetry reflects a decrease in the capacitance value of an AP-aligned junction through which electrons tunnel into the island. This decrease is quantitatively described by spin-dependent drift-diffusion model. In AP configurations, spin-accumulation causes a difference between spin-up and spin-down diffusion lengths that provides a ground for the creation of a tiny charge dipole. This charge dipole acts as an extra serial capacitance that gives rise to the observed TMC effect. The magnetocapacitance also implies an extra energy threshold for a single spin to flip whenever it is entering the island through an AP-aligned tunnel junction.

**References**

1. Ramesh, R. & Spaldin, N. A. Multiferroics: progress and prospects in thin films. *Nature Matl.* **6,** 21–29 (2007).
2. Dong, S., Li, J.-F. & Viehland, D. Ultrahigh magnetic field sensitivity in laminates of TERFENOL-D and Pb ( Mg 1/3 Nb 2/3 ) O 3 – PbTiO 3 crystals. *Appl. Phys. Lett.* **83,** 2265 (2003).
3. Eerenstein, W., Mathur, N. D. & Scott, J. F. Multiferroic and magnetoelectric materials. *Nature* **442,** 759–765 (2006).
4. Kimura, T. *et al.* Magnetocapacitance effect in multiferroic BiMnO_3. *Phys. Rev. B* **67,** 180401 (2003).
5. Mamin, R., Egami, T., Marton, Z. & Migachev, S. Giant dielectric permittivity and magnetocapacitance in La0.875Sr0.125MnO3 single crystals. *Phys. Rev. B* **75,** 115129 (2007).
6. Chang, Y.-M. *et al.* Extraction of the tunnel magnetocapacitance with two-terminal measurements. *Jour. of Appl. Phys.* **107,** 093904 (2010).
7. Kaiju, H., Fujita, S., Morozumi, T. & Shiiki, K. Magnetocapacitance effect of spin tunneling junctions. *Jour. of Appl. Phys.* **91,** 7430 (2002).
8. Padhan, P., LeClair, P., Gupta, A., Tsunekawa, K. & Djayaprawira, D. D. Frequency-dependent magnetoresistance and magnetocapacitance properties of magnetic tunnel junctions with MgO tunnel barrier. *Appl. Phys. Lett.* **90,** 142105 (2007).
9. Zhang, S. Spin-Dependent Surface Screening in Ferromagnets and Magnetic Tunnel Junctions. *Phys. Rev. Lett.* **83,** 640(1–4) (1999).
10. Chui, S. T. & Hu, L. ac transport in ferromagnetic tunnel junctions. *Appl. Phys. Lett.* **80,** 273(1–3) (2002).
11. Son, P. C. V., Kempen, H. & Wyder, P. Boundary Resistance of the Ferromagnetic-Nonferromagnetic Metal Interface. *Phys. Rev. Lett.* **58,** 2271(1–3) (1987).
12. Brataas, A., Nazarov, Y. & et al. Spin accumulation in small ferromagnetic double-barrier junctions. *Phys. Rev. B* **59,** 93-96 (1999).
13. Chui, S. T. Electron interaction on the giant magnetoresistance in the perpendicular

geometry. *Phys. Rev. B* **52,** R3832(1–4) (1995).
14. Hemberger, J. *et al.* Relaxor ferroelectricity and colossal magnetocapacitive coupling in ferromagnetic CdCr2S4. *Nature* **434,** 364–367 (2005).
15. *Single Charge Tunneling* edited by Grabert, H. and Devoret, M. H. Vol. 294, 65-91 (Plenum Press, New York, 1992).
16. Dubois, S. *et al.* Evidence for a short spin diffusion length in permalloy from the giant magnetoresistance of multilayered nanowires. *Phys. Rev. B* **60,** 477–484 (1999).
17. Deutscher, G. & deGennes, P. G. *Superconductivity* edited by Parks, R. D. (Marcel Dekker, 1969).
18. Clarke, J. The Proximity Effect Between Superconducting And Normal Thin Films In Zero Field. *Jour. De Phys.* **29,** C2–3
19. Werthamer, N. R. Theory of the Superconducting Transition Temperature and Energy Gap Function of Superposed Metal Films. *Phys. Rev.* **132,** 2440(1–6) (1963).
20. Eiles, T. M., Martinis, J. M. & Devoret, M. H. Even-Odd Asymmetry of a Superconductor Revealed by the Coulomb Blockade of Andreev Reflection. *Phys. Rev. Lett.* **70,** 1862(1–4) (1993).
21. Ku, H. Y. & Ullman, F. G. Capacitance of Thin Dielectric Structures. *Jour. of Appl. Phys.* **35,** 265(1–3) (1964).

**Table 1** | $\Delta_{TMC}$ in percentage (%) calculated for each of four borders of Coulomb diamonds in the three configurations that exhibit asymmetric diamond. All values are within an error bar of 4%.

| Configuration | $L_{in}$ | $R_{out}$ | $L_{out}$ | $R_{in}$ |
|---|---|---|---|---|
| Left-AP | 2.4 | ***40.3*** | 3.2 | 0.7 |
| Right-AP | -0.9 | 0.6 | ***31.0*** | 2.9 |
| Both-AP | 1.2 | ***41.8*** | ***30.3*** | 1.8 |

**Acknowledgements**

We thank Tzu-Hui Hsu, Tao-Hsiang Chung, and Dr. Min-Chou Lin for their assistance during the course of this work. We are also indebted to Professor Tsan-Chuen Leung and Dr. Jyh-Yang Wang for fruitful discussions. This research was funded by the National Science Council of Taiwan under contract No. NSC 101-2112-M-001-028-MY3. Technical support from NanoCore, the Core Facilities for Nanoscience and Nanotechnology at Academia Sinica, is also acknowledged.

**Author contributions**

T.H.L. fabricated the sample. T.H.L. conducted the experiments. T.H.L. wrote the manuscript. C.D.C. supervised the project. All authors discussed the results, analyzed the data, and reviewed the manuscript.

# Supplementary Information for "Probing Spin Accumulation induced Magnetocapacitance in a Single Electron Transistor"

Teik-Hui Lee[1,2,3] and Chii-Dong Chen[3]

[1]*Department of Physics, National Taiwan University, Taipei 106, Taiwan*

[2]*Nano Science and Technology Program, Taiwan International Graduate Program, Academia Sinica, Taipei 115, Taiwan*

[3]*Institute of Physics, Academia Sinica, Taipei 115, Taiwan*

**S1. Spin-dependent drift-diffusion model:**

At any instance of time $t$ and any location $x$ inside Al layer, we have the spin-dependent electron densities, $n_\pm(x,t)$, and chemical potentials $V_\pm(x,t)$. At the Al/Py interface ($x=x_b$), the spin polarization of Al, $P(x,t) = [n_+(x,t) - n_-(x,t)]/[n_+(x,t) + n_-(x,t)]$ is exchange-induced by the much thicker Py as a result of proximity effect [1,2]. Therefore, the initial value for $P(x,t)$ is non-zero, spatial-independent, and takes up a value close to the one found in Py, i.e., $P_0 \approx P_{Py}$. At $x=x_a$, there is a barrier interface Al$_2$O$_3$/Al where electrons tunnel into Al from Co, giving rise to spin-dependent current density, which can be estimated based on Jullière model:

$$J_\pm(x=x_a) = J \frac{[(1 \pm P_{Co})(1 \pm P_0)]}{[(1+P_{Co})(1+P_0) + (1-P_{Co})(1-P_0)]} \qquad (1)$$

where $P_{Co}$ is spin polarization of Co, and $J=-I/A$

In All-P configuration, we can assume $P_{Co} = P_0$. The spin-dependent but spatial-independent current density is $J_\pm = J(1 \pm P_0)/2$, and thus there is no spin-accumulation. On the other hand, there is charge screening, which in All-P configuration is calculated below. First, the boundary condition of the spin-dependent $E$-field (gradient of $V_\pm(x,t)$) is determined such that $\partial V_+/\partial x = \partial V_-/\partial x = -\Delta V/d$ is a constant at $x=x_a$, while vanishes at $x=x_b$ (Figure S1a). The latter is set under the assumption that the electric field vanishes at the Al/Py interface. We then consider the spin-dependent driven-drift current caused by the boundary $E$-field, i.e., $J_{ext\pm}(x,t) = -\sigma_\pm(x,t)\partial V_\pm(x,t)/\partial x$, with spin-dependent conductivity of Al, $\sigma_\pm(x,t) = 0.5[1 \pm P(x,t)]\sigma_{Al}$, where $\sigma_{Al}$ is the intrinsic conductivity of Al. Plugging this in the continuity equation we obtain spin-dependent depletion causing by external $E$-field after an infinitesimal time $dt$:

$$dn_{ext\pm}(x,t) = -\sigma_\pm(x,t)\frac{\partial^2 V_\pm(x,t)}{e\partial x^2}dt \tag{2}$$

This $dn_{ext\pm}(x,t)$ yields a deviation $\delta_\pm(x,t)dn_{ext\pm}(x,t)$ from equilibrium Fermi level, where $\delta_\pm(x,t) \propto n_\pm^{-1/3}(x,t)$ is the level spacing inside Al. As a result, a steady back-diffusion current is established, i.e., $J_{diff\pm}(x,t) = -\sigma_\pm(x,t)\delta_\pm(x,t)\partial n_\pm(x,t)/\partial x$. Using again the continuity equation, we obtain the refill caused by internal back-diffusion for each spin channel:

$$dn_{diff\pm}(x,t) = D_\pm(x,t)\frac{\partial^2 n'_{m,\pm}(x,t)}{\partial x^2}dt \tag{3}$$

, where $D_\pm(x,t) = \delta_\pm(x,t)\sigma_\pm(x,t)/e \propto n_\pm^{2/3}(x,t)$ is the spin-dependent diffusion coefficient. Summing up contributions from $dn_{ext\pm}(x,t)$ and $dn_{diff\pm}(x,t)$ after each $dt$, the new density distribution for each spin is $n_\pm(x,t+dt) = n_\pm(x,t) - dn_{ext\pm}(x,t) + dn_{diff\pm}(x,t)$, as shown in Figure S1b. This deviation from the initial uniform charge density $n_{\pm,0}$, i.e., $e[n_\pm(x,t) - n_{\pm,0}]$, modifies the chemical potential profile $V_\pm(x,t)$, as described by the spin-dependent Poisson's equation:

$$\frac{\partial^2}{\partial x^2}\left[\frac{1}{2}(1 \pm P(x,t))V_\pm(x,t)\right] = \frac{e[n_\pm(x,t) - n_\pm(x,0)]}{\varepsilon_0} \tag{4}$$

, where $\varepsilon_0$ is free space electric permittivity. At this point $V_\pm(x,t)$ can be solved and plugged back in equation (2) to form a complete iteration loop. We note that, in All-P configuration, even though spin-up and spin-down charge densities are different, but due to nonzero $P(x,t)$, the chemical potential for spin-up, and spin-down are the same at all instants, i.e. $V_+(x,t) = V_-(x,t)$.

In addition, the average chemical potential $V(x,t) = V_+(x,t)[1+P(x,t)]/2 + V_-(x,t)[1-P(x,t)]/2$ is spin-independent. Upon convergence, the sum $dn_+(x,t)/dt + dn_-(x,t)/dt = 0$ is accomplished, giving time-independent $V(x)$ and $n_\pm(x)$ distributions. The total charge density perturbation is then given by

$$\Delta n(x) = n_+(x) + n_-(x) - n_0 \tag{5}$$

Where $n_0 = n_{+,0} + n_{-,0}$. Accordingly, the interfacial capacitance $C_{Al} = e\Delta n(x) A dx / dV(x)$ can be calculated.

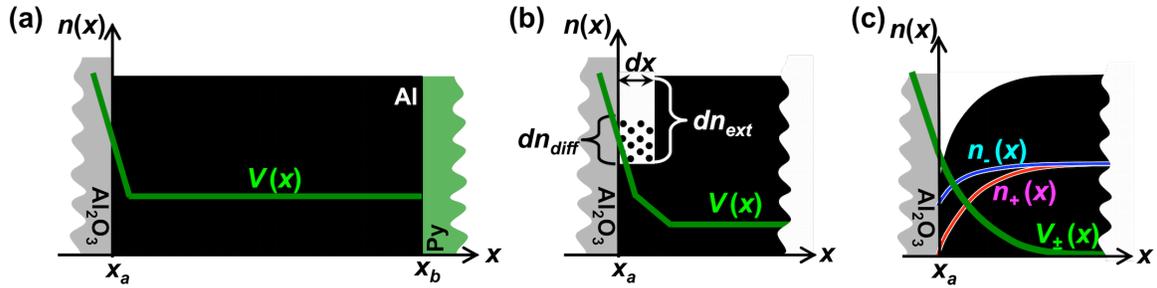

**Figure S1 (color online) | Charge screening process in All-P configuration.** (a) Initial $n(x)$ and $V(x)$ profiles. The dark area shows initial uniform free electron density inside Al and the green line shows initial $V(x)$ at $t=0$. (b) $n(x)$ and $V(x)$ profiles after the first iteration. Scooped area represents depleted density $dn_{ext}(x)$ and dotted area shows refilled back-diffusion density $dn_{diff}(x)$. (c) Saturated $n(x)$ and $V(x)$. Both curves have exponential form with a characteristic length $x$. Red and blue curve inside dark area represent the spin up and down components of $n(x)$, respectively.

Note that both $V(x)$ and $\Delta n(x)$ follow an exponential decay with the screening length $x$, as depicted in Figure S1c, but the calculated $C_{Al}$ is constant everywhere inside Al. This screening length $x$ turns out to be the same as that for the case of non-magnetic junctions[3], in which $P(x,t)=0$, and $n_+(x)=n_-(x)$, $V_+(x)=V_-(x)$. The calculated $C_{Al}$ serves as a baseline for the calculations of extra $C_S$ in the AP-configurations.

**S2. Spin-dependent diffusion lengths**

In AP-configuration, we assume $P_{Co}=-P_0$, and equation (1) is reduced to $J_\pm(x=x_a)=J/2$ (Figure 3a in *Main Text*). Consequently, the spin dependent current density obeys diffusion equation, which in steady state is given as:

$$dJ_\pm(x,t) = D_\pm(x,t)\left[\frac{\partial^2(J_\pm(x,t)-J_\pm(x,0))}{\partial x^2} - \frac{J_\pm(x,t)-J_\pm(x,0)}{\lambda_\pm^2}\right]dt \qquad (6)$$

After each $dt$, we have $J_\pm(x,t+dt)=J_\pm(x,t)+dJ_\pm(x,t)$, and the accumulated/depleted spin density is [4-6]:

$$dn_{acc,\pm}(x,t) = \frac{\partial J_\pm(x,t)}{e\partial x}dt \qquad (7)$$

This spin-accumulation causes a difference in the chemical potential for spin-up and spin-down electrons. This difference in turn triggers spin flips and that produces a "refill" of spin density:

$$dn_{flip,\pm}(x,t) = \left[3(1\mp P(x,t))D_\pm(x,t)\frac{n_\pm(x,t)\delta_\pm(x,t)-n_\mp(x,t)\delta_\mp(x,t)}{4\delta_\pm(x,t)\lambda_\pm^2}\right]dt \qquad (8)$$

, where $n_\pm(x,t)\delta_\pm(x,t) - n_\mp(x,t)\delta_\mp(x,t)$ is the chemical potential difference. In the mean time, spin accumulation also cause a difference in diffusion coefficient for each spin, $D_\pm(x,t) \propto n_\pm^{2/3}(x,t)$, which depends on their respective chemical potential. As a result, each spin has a different diffusion length, since $\lambda_\pm \propto \sqrt{D_\pm}$. Following Eq. (2) and (3), the total spin density at $t=t+dt$ is the sum of all contributions, *i.e.*,

$$n_\pm(x,t+dt) = n_\pm(x,t) - dn_{ext,\pm}(x,t) + dn_{diff,\pm}(x,t) + dn_{acc,\pm}(x,t) - dn_{flip,\pm}(x,t) \quad (9)$$

This is then plugged back in Eq. (4) to complete the new iteration loop. Finally, upon convergence, the system reaches a steady state so that the sums $dn_\pm(x,t)/dt - dn_\mp(x,t)/dt = 0$ and $dn_+(x,t)/dt + dn_-(x,t)/dt = 0$ give the new time-independent $V_\pm(x)$ and $n_\pm(x)$ distributions. By updating Eq. (5), we deduced $C_S$ in AP-configuration (in the *Main Text*).

**S3. Formation of Charge dipole by Spin Accumulation**

Our device is composed of tunnel junctions that comprise of structure as follow: Co/Al$_2$O$_3$/Al/Py. Since Al is located between magnetically aligned Co and Py, it serves as an ideal place to conceive spin accumulation in AP configuration. The current flowing through tunnel barrier obeys Julliere's model as in Eq. (1). Hence in AP configuration, current for spin up and spin down are in same proportion to each other. Since Al is exchange induced in Ohmic proximity by Py, the current inside Al layer is initially uniformly polarized. As a result, for each spin, there is a discrepancy between tunneling current at Al$_2$O$_3$/Al interface and the current inside Al. As in Eq. (6), due to this discrepancy, the current for each spin diffuse away from the Al$_2$O$_3$/Al interface towards Py. This non-steady of the flowing current, manifested as the gradient of the current density for each spin as shown in Eq. (7), causing each spin to either accumulate or deplete depending on the flowing direction. Spin-down (blue) is accumulated when electrons flow from Al$_2$O$_3$/Al interface towards Py (Figure S2a), while spin-up (red) is accumulated when electrons flow otherwise (Figure S2b). As electron density determine diffusion coefficient, thus the spin component that is accumulating has a longer diffusion length as compared to the one that is depleting. And the total accumulated spin across Al must equate the total depleted spin, due to the conservation of charge.

$$\int_{x_a}^{x_b} \Delta n_{S,+}(x)\, dx = \int_{x_a}^{x_b} \Delta n_{S,-}(x)\, dx \tag{10}$$

Where $\Delta n_{S,\pm}(x) = \int_0^{\infty} \frac{\partial J_{\pm}(x,t)}{e \partial x} dt$ is the accumulated/depleted amount of spin due to non-steady current of each spin. Consequently, the accumulating spin is less than the depleting spin at Al$_2$O$_3$/Al interface, while otherwise at Al/Py interface, thus forming a charge dipole within Al (Figure S2c). Note that the direction of the created charge dipole is independent of the electron flowing direction.

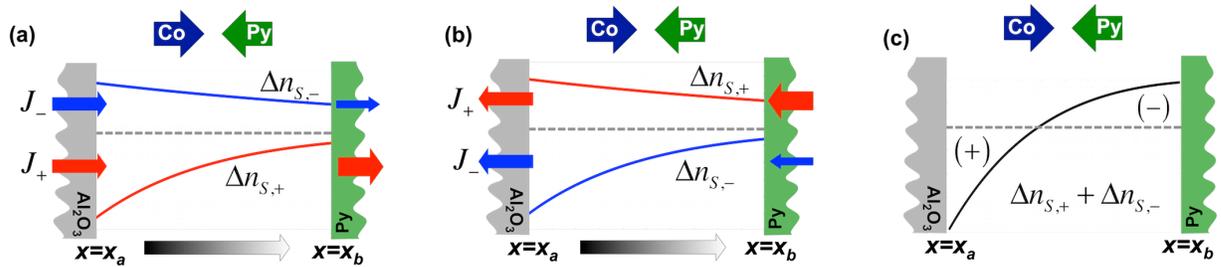

**Figure S2 (color online) | Charge dipole formation inside Al layer due to spin accumulation.** Grey line indicates intrinsic uniform electron density. Note that diffusion (gradient grey arrow at bottom) always diffuse away from tunnel barrier. (a) Electrons flow from left to right. Current density $J$ is composed of spin up (red) and spin down (blue) components, represented by corresponding colored arrows, whose width indicates the relative magnitude. (b) Electrons flow from right to left. (c) Charge dipole as the summation of spin-up and spin-down electron densities

## S4. Numeric calculation of $\Delta_{TMC}$ value

Within our model, the $\Delta_{TMC}$ value increases with the spin diffusion length $\lambda$. For $\lambda \to 0$, there is no spin accumulation and the All-P case is recovered so that $\Delta_{TMC} \to 0$. Contrarily, in the case of $\lambda \to \infty$, it is not possible to flip spins, and the energy required to flip a single spin ($e^2/2C_S$ in Fig. 3c) becomes infinite so that $\Delta_{TMC} \to 100\%$. In the numeric calculations, due to some technique issues (such as limited memory and CPU speed), we could only get a $\Delta_{TMC}$ value of up to 80%. The calculated $\Delta_{TMC}$ as a function of spin diffusion length $\lambda$ using the experimental device parameters and the reported polarization values[7] for Co (35%) and Py (40%) is shown in Fig. S3. It is found that, for a given $\lambda$, the $\Delta_{TMC}$ value is insensitive to the magnitude of tunneling current and the polarization values. However, there are factors such as interfacial roughness and impurities in Al that are not considered in the present calculation. For the observed $\Delta_{TMC} \approx 40\%$, this calculation implies a spin diffusion length of about 2.8nm, and the dipole location $x_c$, which is related to the spin diffusion length, is calculated to be about 1.2nm.

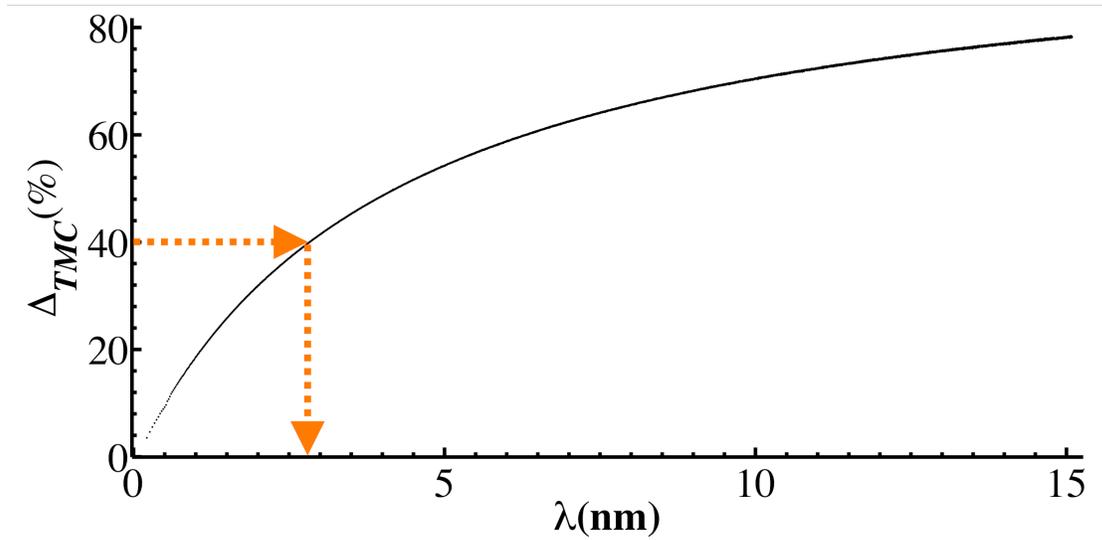

**Figure S3 (color online) | Calculated $\Delta_{TMC}$ value as a function of spin diffusion length $\lambda$ in Al layer under AP-configuration**. Orange dash arrow corresponds to the experimentally observed $\Delta_{TMC}$, suggesting a spin diffusion length of 2.8nm.

**References**

1. Mitsuzuka, T., Matsuda, K., Kamijo, A. & Tsuge, H. Interface structures and magnetoresistance in magnetic tunnel junctions. *Journal of Applied Physics* **85,** 5807 (1999).
2. Moodera, J., Taylor, M. & Meservey, R. Exchange-induced spin polarization of conduction electrons in paramagnetic metals. *Physical Review B* **40,** 11980–11982 (1989).
3. Ku, H. Y. & Ullman, F. G. Capacitance of Thin Dielectric Structures. *Journal of Applied Physics* **35,** 265(1–3) (1964).
4. Johnson, M. Spin accumulation in gold films. *Physical Review Letters* **70,** 2142 (1993).
5. Zaffalon, M. & van Wees, B. Spin injection, accumulation, and precession in a mesoscopic nonmagnetic metal island. *Physical Review B* **71,** 125401 (2005).
6. Jedema, F., Nijboer, M. & et al. Spin injection and spin accumulation in all-metal mesoscopic spin valves. *Physical Review B* **67,** 085319 (2003).
7. Meservey, R. & Tedrow, P. M. Spin-polarized electron tunneling. *Physics Reports* **238,** 173–243 (1994).